\documentclass[twocolumn,pre,aps,showpacs,amsmath,amssymb,superscriptaddress]{revtex4}

\usepackage{hyperref}
\usepackage{amsmath,amssymb}
\usepackage{amsfonts,amsthm}
\usepackage{graphics}
\usepackage{graphicx}
\usepackage{dcolumn}
\usepackage{color}
\usepackage{bm}

\usepackage[normalem]{ulem}

\begin{document}

\title{Inhibitory loop robustly induces anticipated synchronization in neuronal microcircuits}
\author{Fernanda S. Matias}
\thanks{fernanda@fis.ufal.br}
\affiliation{Instituto de F\'{\i}sica, Universidade Federal de Alagoas, Macei\'{o}, Alagoas 57072-970 Brazil}
\author{Leonardo L. Gollo}
\affiliation{System Neuroscience Group, Queensland Institute of Medical Research, Brisbane QLD 4006, Australia}
\author{Pedro V. Carelli}
\affiliation{Departamento de F\'{\i}sica, Universidade Federal de Pernambuco, Recife, Pernambuco 50670-901 Brazil}
\author{Claudio R. Mirasso}
\affiliation{Instituto de Fisica Interdisciplinar y Sistemas Complejos, CSIC-UIB, Campus Universitat de les Illes Balears E-07122 Palma de Mallorca, Spain}
\author{Mauro Copelli}
\affiliation{Departamento de F\'{\i}sica, Universidade Federal de Pernambuco, Recife, Pernambuco 50670-901 Brazil}

\begin{abstract}

We investigate the synchronization properties between two excitatory coupled neurons 
in the presence of an inhibitory loop mediated by an interneuron.
Dynamical inhibition together with noise independently applied to each neuron provide phase diversity in the dynamics of the neuronal motif.
We show that the interplay between the coupling strengths and external noise
controls the phase relations between the neurons in a counter-intuitive way.
For a master-slave configuration (unidirectional coupling) we find that the slave can anticipate the master,
on average, if the slave is subject to the inhibitory feedback.
In this non-usual regime, called anticipated synchronization (AS), 
the phase of the post-synaptic neuron is advanced with respect to that of the pre-synaptic neuron.
We also show that the AS regime survives even in the presence of unbalanced bidirectional excitatory coupling.
Moreover, for the symmetric mutually coupled situation, the neuron that is subject to the inhibitory loop leads in phase.
\end{abstract}
\maketitle

\section{Introduction}
%
%

Neuronal synchronization is a common feature of nervous systems~\cite{Wang10}.
According to the principle of communication through  coherence~\cite{Fries05}, 
the phase difference between sender and receiver 
circuits influences the effectiveness of the information transmission~\cite{Bastos15,Tiesinga10}.
Recent studies showing non-zero phase lag between synchronized areas in the brain~\cite{Dotson14,Maris13,Jia13,Grothe12,Liebe12,Phillips14}
have sparked interest in the potential function of the phase diversity~\cite{Maris16}.
In electrical brain signals the phase was usually associated to delays 
in axonal transmission and synaptic effects~\cite{Marsden01,Williams02,Schnitzler05,Sauseng08,Gregoriou09}.
However, modeling studies have shown that the phase difference 
can be determined, among other things, by a local inhibitory loop at the receiving end~\cite{Matias11,Matias14}, a mechanism which could explain 
unexpected negative phase lags found in neuronal data~\cite{Matias14,Brovelli04,Salazar12}.

The counterintuitive kind of synchronization in which a unidirectionally coupled system 
exhibits negative phase lag is called anticipated synchronization (AS)~\cite{Voss00}. 
AS, as proposed by Voss~\cite{Voss00}, was originally defined between two identical autonomous dynamical systems coupled in an 
unidirectional (``master-slave'') configuration  in the presence of a negative 
delayed feedback. Such system is described by the following equations:
\begin{eqnarray}
\label{eq:voss}
\dot{\bf {m}} & = & {\bf f}({\bf m}(t)), \\
\dot{\bf {s}} & = & {\bf f}({\bf s}(t)) + K[{\bf m}(t)-{\bf s}(t-t_d)]. \nonumber 
\end{eqnarray}
${\bf m}$ and ${\bf s}$ $\in \mathbb{R}^n$ are dynamical 
variables respectively representing the master and the slave systems, 
${\bf f}$ is a vector function which defines 
each autonomous dynamical system, $K$ is a coupling matrix and $t_d > 0$ is the delay in the slave's negative self-feedback. 
In such system, 
${\bf s}(t) = {\bf m}(t+t_d)$ is a solution of the system, 
which can be easily verified by direct substitution 
in Eq.~\ref{eq:voss}. The striking aspect of this 
solution is its meaning: the state of the receiver
system ${\bf s}$ anticipates the sender's state ${\bf m}$. 
In other words, the slave predicts the master behavior. 

After several 
theoretical~\cite{Masoller01,HernandezGarcia02,Calvo04,Kostur05,Wang05a,Ambika09,Senthilkumar05,Pyragiene15,Ciszak15} 
and experimental~\cite{Sivaprakasam01,Ciszak09,Pisarchik08,Blakely08,Liu02,Tang03,Corron05}
works in physical systems, a reasonable question was whether AS could occur in natural (not man-made) systems. 
The first verification of AS in a neuronal model was done by Ciszak et al.~\cite{Ciszak03} using two 
unidirectionally and electrically coupled FitzHugh-Nagumo neuron models 
in the presence of a negative delayed self-feedback in the slave.
Though potentially interesting for neuroscience, it is not trivial to compare these theoretical results with real neuronal data. 
It is not evident how (or whether) the delayed inhibitory self-coupling 
of the slave membrane potential employed by Ciszak et al.~\cite{Ciszak03,Ciszak04,Ciszak09} could be implemented 
by the brain.

Recently, AS was found in more realistic neuronal models such as nonidentical chaotic neurons~\cite{Pyragiene13}, map-based 
neurons with a memory term~\cite{Sausedo14} and two Hodgkin-Huxley neurons with different depolarization parameters~\cite{Simonov14}.
In particular, it has been shown that periodically spiking neurons can show AS within a plausible biological scenario, in which
the delayed self-feedback of Eq.~(\ref{eq:voss}) was replaced by an inhibitory loop mediated by chemical synapses~\cite{Matias11}. 
Those results were later extended to assess the effects of spike-timing-dependent plasticity~\cite{Matias15}.
Furthermore, AS mediated by inhibition has also been found in a model of neuronal populations 
which can explain coherent oscillations with the negative phase-lag observed between areas of the monkey cortex~\cite{Matias14}.

Here we study the phase diversity induced by anticipated synchronization due to
dynamical inhibition and noise in a neuronal motif described by the Hodgkin-Huxley equations, which is introduced in section~\ref{model}. 
In section~\ref{results} we present our results, showing that
in a more realistic scenario where neurons are subject to independent noise realizations,
the anticipation does not occur for every spike, but survives on average.
Moreover, we find that the mean spike-timing difference between master and slave neurons (see Fig.~\ref{fig:MSI}) is a function of 
the inhibitory conductance, which controls the phase diversity.
We also show that AS is robust in the presence of an excitatory feedback from the slave 
to the master. Finally,
in section~\ref{conclusions} we present our conclusions and
briefly discuss the potential significance of our results for neuronal circuits.

\section{\label{model}Model}

\begin{figure}
\centering
\includegraphics[width=0.8\columnwidth,clip]{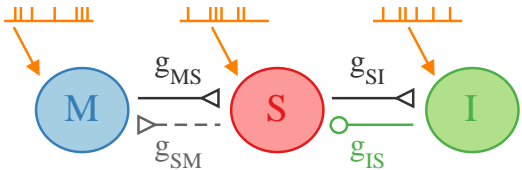}
\caption{
{\bf (Color online) Neuronal motif: master-slave-interneuron (MSI)}
Each node represents a neuron described by the Hodgkin-Huxley model connected to other neurons by chemical synapses.
The important parameters to determine the mean spike-timing difference $\tau$ between M and S are the synaptic conductances $g_{xy}$ 
(where $x,y=$M,S,I indicate the pre-synaptic and the post-synaptic neurons, respectively).
Each neuron is subject to an independent Poisson spike train.
}
\label{fig:MSI}
\end{figure}

The results obtained in Ref.~\cite{Matias11} are important as a first step towards the demonstration that AS 
can indeed occur in neuronal circuits. 
However, the assumptions of strict unidirectionality and absence of noise call into question the robustness of the results.
In this paper we study new configurations and assume stochastic input currents to verify whether AS holds in a more realistic environment.  
In our scheme, sketched in Fig.~\ref{fig:MSI}, the two excitatory neurons, master (M) and slave (S), are bidirectionally connected 
via excitatory chemical synapses (with synaptic conductance $g_{MS}$ and $g_{SM}$).
The slave neuron feeds an interneuron $I$ with conductance $g_{SI}$. The interneuron feeds S back with an inhibitory chemical synapse with conductance $g_{IS}$.
The three neurons M, S and I are subject to an independent Poisson spike train.
This kind of configuration is found in several circuits including the spinal cord~\cite{Shepherd}, the
thalamus~\cite{thalamus1,thalamus2} the olfactory system~\cite{Kay06,Rospars14}, 
the motor circuits for self-regulation~\cite{Kandel}. 
It was also proposed in hybrid experiments with real and simulated neurons~\cite{Masson02}.
In particular this motif is one of the most over-represented 3-neuron motifs of the \textit{C. elegans} connectome~\cite{Qian11}.
We will refer to this circuit as the MSI motif. 
Each neuron here is modeled by the Hodgkin-Huxley (HH) equations~\cite{Koch,HH52}, 
whereas chemical synapses are modeled with standard first-order kinetic equations~\cite{KochSegev}. 
Details of the model are described in the Appendix.

In the HH model, the external constant current $I_c$ determines the activity of the neuron
when all other currents are zero~\cite{Rinzel80}.
For the chosen parameters and $I_c \lesssim 177.13$~pA the neuron is in a stable fixed point. 
For $177.13$~pA~$\lesssim I_c \lesssim 276.51$~pA, the stable fixed point coexists with a stable limit cycle.
For $I_c \gtrsim 276.51$~pA  the fixed point loses stability (via a subcritical Hopf bifurcation) and the neuron spikes periodically with a frequency that  increases slightly with $I_c$. 
Unless otherwise stated, each neuron is in the excitable regime subject to an external constant current $I_c=170$~pA 
and to an independent noisy spike train described by a Poisson distribution with rate $R$ (see Appendix for details).

\begin{figure}
\centering
\includegraphics[width=0.98\columnwidth,clip]{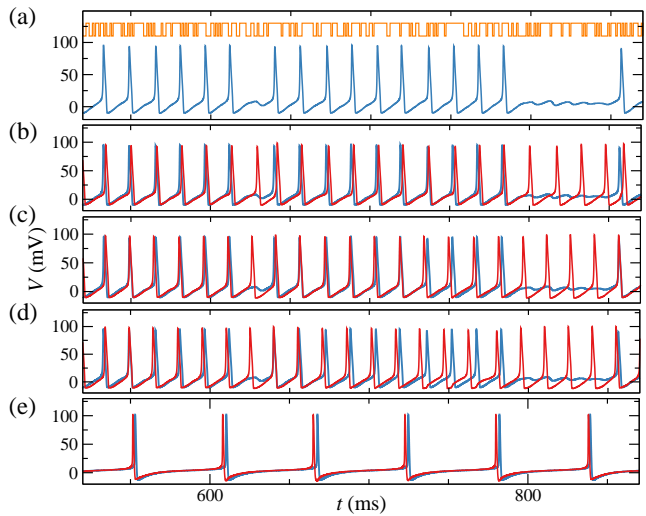}
\caption{
{\bf (Color online) Membrane potential of the neurons in the presence of noise.}
(a) Time series of the master neuron receiving both a constant external current $I_c=170$~pA 
and an excitatory input, obeying the Poisson statistic shown on top, mediated by chemical synapses with neurotransmitter concentration $[T]$.
(b)-(d) Time traces of the slave neuron for different values of the inhibitory conductance $g_{IS}$.
As we increase the inhibition, the order between M and S spikes change from pre-post (MS) to post-pre (SM):
(b) $g_{IS}=10$~nS (DS),
(c) $g_{IS}=40$~nS (AS),
(d) $g_{IS}=60$~nS (AS). 
(e) Time series of M and S neurons with lower frequencies, modeled by the modified Hodgkin-Huxley equations 
described in the appendix with $g_{MS}=50$~nS and $g_{IS}=150$~nS. Note that to represent all data in the same scale we have added +55 mV to the membrane potential in the  modified model (see appendix for details).
}
\label{fig:Vmemb}
\end{figure}

An example of the evolution of the neuronal membrane potential for an external current 
$I_c=170$~pA and a Poisson input with rate $R=63$~Hz is shown in Fig.~\ref{fig:Vmemb}a. 
The neurotransmitter concentration $[T]$ is represented by the Poisson train at the top of panel (a).
Note that the noise effectively puts the neuron in the bi-stable regime. 
Time series for the master and slave neurons are shown in Fig.~\ref{fig:Vmemb}b-d for different $g_{IS}$.
The S and I neurons also receive \textit{different, independent} realizations of the Poisson train, with the same rate, which are not shown.
As can be seen, inhibition affects the time difference between the spikes of the master and the slave in each cycle.
To probe the generality of the phenomenon at lower frequencies, we used a modified version of the Hodgkin-Huxley model
that contains an extra delayed-rectifier slow  K$^+$ current (see appendix for details). As it is shown in Fig.~\ref{fig:Vmemb}e, AS is also observed for these slower pulsating neurons. In the following, and without loss of generality, we concentrate in the standard Hodgkin-Huxley model.

\section{\label{results}Results} 

\subsection{\label{sec:MSI} The inhibitory loop entails a counterintuitive average spike-timing difference}

\begin{figure}
\begin{minipage}{0.98\linewidth}
\centering
\includegraphics[width=0.9\columnwidth,clip]{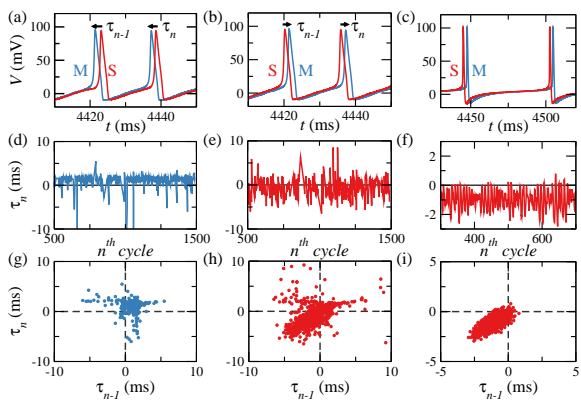}
\end{minipage}
\caption{
{\bf (Color online) Characterization of the delayed synchronization regime (DS) and the anticipated synchronization regime (AS).} 
Left panels $g_{IS}=10$~nS (DS, $\tau>0$), middle panels $g_{IS}=40$~nS (AS, $\tau<0$), right panels modified Hodgkin-Huxley model (AS, $\tau<0$, $g_{MS}=50$~nS and $g_{IS}=150$~nS).
(a)-(c) The spike-timing differences ($\tau_{n-1}$ and $\tau_n$) between a spike of the master neuron 
and a spike of the slave neuron in two consecutive periods.
(d)-(f) $\tau_n$ as a function of the cycle number $n$.
(g)-(i) Return map  $\tau_n$ \textit{versus} $\tau_{n-1}$.
}
\label{fig:3x3}
\end{figure}

Differently from the system described by Eq.~\ref{eq:voss}, when we replace the self-feedback loop by a dynamical inhibition,
the spike-timing difference between the master and the slave neurons is not hardwired anymore. 
It emerges as a property of the system dynamics, depending on the synaptic parameters~\cite{Matias11}.
We define the spike-timing difference $\tau_n$ between the M and S neuron in each cycle as the difference:
\begin{equation}
\tau_n \equiv t^{S}_{n}-t^{M}_{n},
\end{equation}
where $t^{M}_{n}$ and $t^{S}_{n}$ are the closest spike times (defined by a peak in the membrane potential) 
of the M and S neurons in each cycle (Fig.~\ref{fig:3x3}).
The spike-timing difference is calculated only if the M neuron produces a spike.
Since we expect $\tau_n$ to be different in each period, we define the average spike-timing $\tau$ 
as the mean value of $\tau_n$ and represent its standard error of the mean as error bars.
Initial conditions were randomly chosen. 
We compute $\tau$ and its standard error of the mean over 
$40$~s of the time trace.
A transient time is always discarded in the simulations.

From our definition, if $\tau_n>0$ the M neuron fires a spike before the S neuron in a given cycle.
As an example, we show in Fig.~\ref{fig:Vmemb}b time series for M and S neurons, subject 
to an independent noise input and synaptic conductances $g_{SM}=0$ and $g_{MS}=g_{SI}=g_{IS}=10$~nS.
In this example, the M neuron (blue) fires before the S neuron (red) in all cycles shown.  
Despite the variations in the values of $\tau_n$
the average spike-timing characterizing this situation is positive: $\tau>0$.

When the M neuron lags behind the S neuron we have $\tau_n<0$.
For example, for the neuronal activity shown in Fig.~\ref{fig:Vmemb}c with the parameters 
$g_{SM}=0$, $g_{MS}=g_{SI}=10$~nS, and $g_{IS}=40$~nS, the S neuron  anticipates the M neuron in at least 12 cycles. 
In this example, the average spike-timing is negative ($\tau<0$) 
even though there are some positive values of $\tau_n$ (see the first cycle, for example).
A similar behavior is found for larger inhibition, for example $g_{IS}=60$~nS in Fig.~\ref{fig:Vmemb}d.
Therefore, the sign of the spike-timing difference $\tau$ determines the neuron that leads the dynamics in each cycle or, in other words, the sign of the relative phase locking.
When $\tau>0$ the activity of M leads that of S, on average.
On the other hand, if $\tau<0$ the S neuron leads the M neuron.
For unidirectional coupling ($g_{SM}=0$ in Fig.~\ref{fig:MSI}) 
only the M neuron sends information to the S neuron
and the nomenclature master-slave is justified. 
The master neuron is the sender and the slave neuron is the receiver.
In this situation, we refer to as delayed synchronization (DS), the regime in which the master leads the slave ($\tau>0$, Fig.~\ref{fig:Vmemb}b), and
anticipated synchronization (AS), the counterintuitive situation 
in which the slave anticipates or ``predicts'' the activity of the master ($\tau<0$, Fig.~\ref{fig:Vmemb}c-d).
This means that the neuron which sends the information lags behind the neuron that receives the information.
Naturally, since the system is in a phase-locked regime, there is no violation of causality, 
nor any real anticipation, the slave's dynamics in any cycle is influenced by the master's dynamics in the preceding cycle(s). 

Another useful way to characterize the synchronization regime is to plot
$\tau_n$ in each cycle (Fig.~\ref{fig:3x3}). 
As we increase $g_{IS}$, the standard error of the mean of $\tau$ also increases, but the average value decreases, reaching negative values.
The return map of the spike-timing difference can also be employed to visualize the AS and DS regimes. 
In the $\tau_{n}$ vs. $\tau_{n-1}$ plane, a larger concentration of points in the first quadrant indicates a DS regime,
whereas a denser region in the third quadrant indicates an AS regime.

\begin{figure}[!ht]
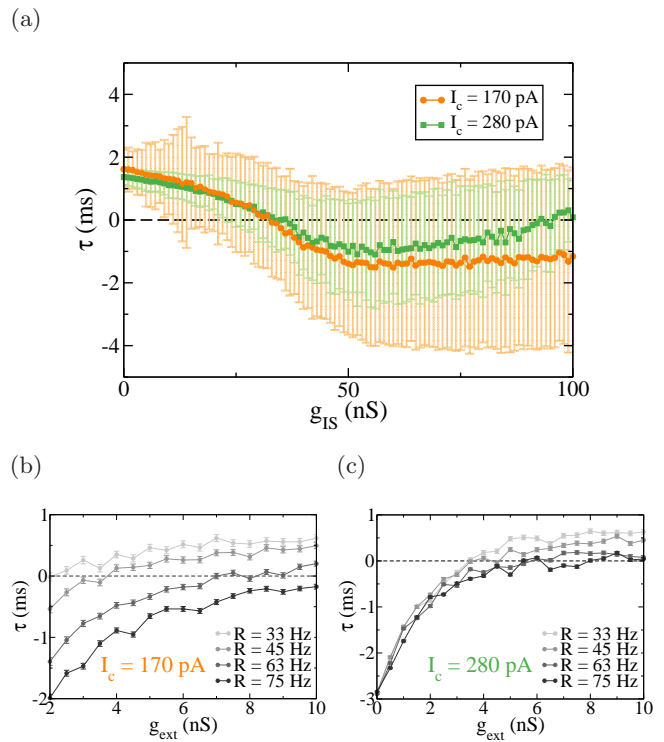
%
\begin{minipage}{0.99\linewidth}
\begin{flushleft}(a)%
\end{flushleft}%
\centerline{\includegraphics[width=0.8\columnwidth,clip]{Matias04a}}
\end{minipage}
\begin{minipage}{0.49\linewidth}
\vspace{0.3cm}
\begin{flushleft}(b)%
\end{flushleft}%
\centerline{\includegraphics[width=0.99\columnwidth,clip]{Matias04b}}
\end{minipage}
\begin{minipage}{0.49\linewidth}
\vspace{0.3cm}
\begin{flushleft}(c)%
\end{flushleft}%
\centerline{\includegraphics[width=0.99\columnwidth,clip]{Matias04c}}
\end{minipage}
\caption{
{\bf (Color online) Comparison of the effects of the inhibition in two different situations:} 
when neurons are in the excitable regime ($I_c=170$~pA) and when neurons are tonically spiking ($I_c=280$~pA). 
(a) The spike-timing difference $\tau$ is a smooth function of the inhibitory conductance $g_{IS}$, which controls the transition from DS to AS 
(with $g_{ext}=2.0$~nS and $R=63$~Hz). Points are the mean $\tau$ and error bars are the standard deviation. The standard error of the mean is comparable to the size of the symbols. 
(b) and (c) $\tau$ as a function of the conductance of the external synapses $g_{ext}$ for different Poissonian rates $R$ and $g_{IS}=50$~nS.
In the excitable regime, if $R$ or $g_{ext}$ are too small the master does not fire a spike. The error bars are the standard error of the mean.
}
\label{fig:tau_gis}
\end{figure}

The average spike-timing difference is a smooth function of the inhibitory conductance (Fig.~\ref{fig:tau_gis}a).
As we increase $g_{IS}$, the system undergoes a continuous transition from DS to AS.
Since the HH model exhibits a Hopf bifurcation, one question that arises is whether 
the existence of the AS regime depends on the applied constant current. 
The answer is that for sufficiently intense noise,
ensuring that the master neuron can fire few consecutive spikes before returning to the silent state, there is always a transition from DS to AS.
In Fig.~\ref{fig:tau_gis}a we plot $\tau$ as a function of  $g_{IS}$ for $I_c=170$~pA and $I_c=280$~pA.
Results are qualitatively similar for intermediate values of $I_c$.
Small values of $I_c$ yield large average spiking-time differences (for both anticipation and delay).
For sufficiently large $I_c$ a second transition from AS to DS for $g_{IS}>90$~nS exists.
The error bars represent the standard deviation $\sigma$. 

In Fig.~\ref{fig:tau_gis} (b) and (c) we investigate the role of the external noise in a systematic way. 
We show how the spike-timing difference changes with both the conductance of the external synapses $g_{ext}$ and the Poissonian rate $R$. 
We plot $\tau$ \textit{versus} $g_{ext}$, for $g_{IS}=50$~nS and $R=33,45,63,75$~Hz.
For $I_c=170$~pA (below the Hopf bifurcation, Fig.~\ref{fig:tau_gis}b), the noise is necessary for the neurons
to fire, whereas for $I_c=280$~pA (beyond the Hopf bifurcation, Fig.~\ref{fig:tau_gis}c) the noise acts as a perturbation.
In both cases, the anticipation time increases with $R$ and decreases with $g_{ext}$.
The error bars represent the standard error of the mean (SEM). 

Finally, in order to assess to which extent different noise sources affect the spread in the spike-timing difference, we also performed simulations with identical Poisson trains  impinging on the M and S neurons. The results are very similar to the ones shown in Fig.~\ref{fig:tau_gis} (a), only with slightly smaller variance. 

\subsection{\label{sec:diversity} Phase relation diversity is modulated by inhibition}

\begin{figure}
\begin{minipage}{0.98\linewidth}
\centering
\includegraphics[width=0.99\columnwidth,clip]{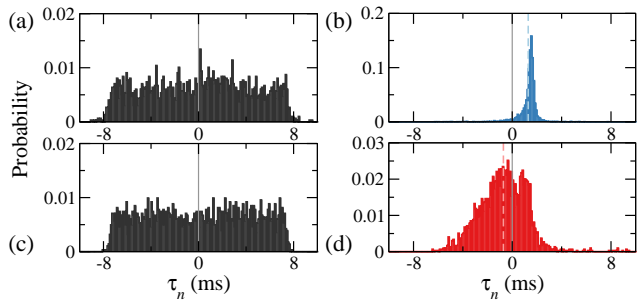}
\end{minipage}
\caption{
{\bf (Color online) Phase diversity for a unidirectional excitatory synapse between master and slave.} 
Histogram of the spike-timing difference between M and S in the uncoupled situation ($g_{MS}=0$, a and c)
and in the MSI motif ($g_{MS}=10$~nS, b and d) for two different inhibitory conductances $g_{IS}=10$~nS (a and b) and
 $g_{IS}=40$~nS (c and d). 
 The dashed lines in b (DS) and in d (AS) indicate that the mean of these distributions are not zero.
 On the contrary the distributions have an almost zero mean in panels a and c.}
\label{fig:histogram}
\end{figure}

Phase relations are considered to play an important role in fast neuronal mechanisms
that underlie cognitive functions~\cite{Maris16}.
For a given frequency band, synchronization is defined as a consistent phase relation between given pairs of neurons 
(in these cases, the relative phase is mapped to the spike-timing difference). 
Non-zero-lag phase differences have been reported in different experiments between individual neurons ~\cite{Maris13,Jia13,Livingstone96,Bastos15}
and between local field potentials measured in different electrodes~\cite{Maris13,Dotson14,Bastos15}. 
In most of them, the phase relation exhibits diversity which, as we show in Fig.~\ref{fig:histogram}, our simple motif model can reproduce. 

In the uncoupled situation ($g_{MS}=g_{SM}=0$), the master and 
the slave-interneuron systems oscillate with similar mean firing rates due to the external input. 
However, the distribution of spike-timing difference between the master and the slave 
is almost uniform (Fig.~\ref{fig:histogram}a), which means that there is neither consistency 
in the phase relation nor synchronization between the neurons.
On the other hand, the MSI motif ($g_{MS}\neq0$) exhibits a richer histogram. 
For weak inhibition, when the neurons fire in the DS regime, there is a sharp unimodal distribution (Fig.~\ref{fig:histogram}b)
at positive $\tau_n$ values.
For stronger inhibition, in the AS regime, the distribution exhibits two peaks: one close to the average spike-timing difference
and a smaller one close to the characteristic time of the excitatory synapse (Fig.~\ref{fig:histogram}d).
Despite the large standard deviation, this histogram is clearly different from the one for the uncoupled case (Fig.~\ref{fig:histogram}c).
We observed that inhibition alone cannot account for the spike-timing difference (Fig.~\ref{fig:histogram}c), 
which depends rather on the interplay between excitation and inhibition impinging on the slave neuron.

\subsection{\label{sec:exc} The excitatory neuron participating in the inhibitory loop leads the other}

For the sake of simplicity, we will maintain the terminology 
master-slave even in the presence of an excitatory feedback
from the slave to the master ($g_{SM}\neq0$, i.e. for mutual coupling).
However, in this situation one should be careful in the determination of the synchronization regimes.
DS refers to the regime in which the sender leads the receiver,
whereas AS refers to the regime in which the receiver leads the sender.
If $g_{MS}> g_{SM}$, the master neuron is the sender (as in our unidirectional situation).
However, if $g_{SM}> g_{MS}$, the slave neuron is the sender.
We aim at understanding how the system behaves as it changes  
from the unidirectional coupling to the completely symmetrical bidirectional coupling.

\begin{figure}
\begin{minipage}{0.98\linewidth}
\centering
\includegraphics[width=0.9\columnwidth,clip]{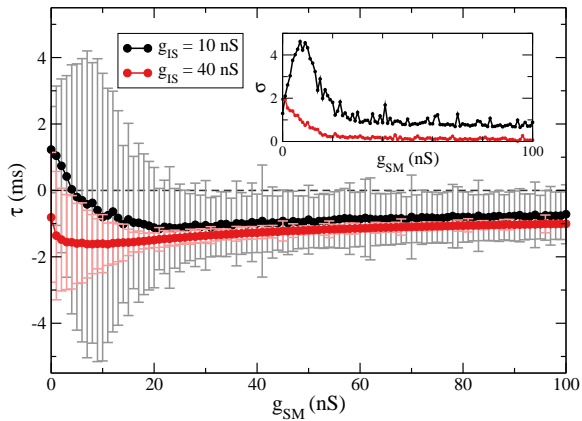}
\end{minipage}
\caption{
{\bf (Color online) Effect of the excitatory feedback from the slave to the master $g_{SM}>0$.} 
The presence of $g_{SM}$  facilitates the leadership of S, which is represented by negative $\tau$. 
Points are the mean $\tau$ and error bars are the standard deviation.
Inset: the standard deviation $\sigma$ as a function of $g_{SM}$. 
In both plots, $g_{MS}=10$~nS.
}
\label{fig:tau_gsm}
\end{figure}

Motivated by the existence of plenty of excitatory neurons that are bidirectionally connected in the brain~\cite{Sporns04}
we analyze the phase effects 
by increasing, from zero, the synaptic conductance $g_{SM}$ in Fig.~\ref{fig:MSI}.
When the system starts in the AS regime for $g_{SM}=0$, increasing 
the excitatory feedback $g_{SM}$ does not change the sign of $\tau$ (the spike-timing difference remains negative, Fig.~\ref{fig:tau_gsm}).
This means that S leads M despite the excitatory feedback. For $g_{SM} \gtrsim g_{MS}$, this is not surprising, since S becomes the effective sender in the microcircuit, whereas M becomes the receiver.

Perhaps less intuitive is the situation where the system is in the DS regime in the absence of excitatory feedback ($g_{IS}=10$~nS in Fig.~\ref{fig:tau_gsm}). 
Increasing $g_{SM}$, a transition from DS to AS occurs even for $g_{SM} < g_{MS}$. 
The transition can be understood as a change of dominance between two competing mechanisms. 
On one extreme, if two identical neurons are unidirectionally connected, phase locking occurs with $\tau > 0$~\cite{Schultheiss}.
On the other extreme, if excitatory neurons with different natural frequencies are mutually connected, the neuron with the largest natural frequency is the leader in a phase locking regime, as can be demonstrated by a simple model of two phase oscillators~\cite{Strogatz}. 
Increasing $g_{SM}$ from zero provides a transition between these two extremes, because in the AS regime the SI subsystem has a higher natural frequency than M (as we have checked numerically). 
In our case, the feedback inhibition in the slave neuron facilitates the slave-interneuron circuit to fire faster than the 
master in the uncoupled situation. For instance, for the parameters used in Fig.~\ref{fig:histogram}c (uncoupled case, $g_{IS}=40$~nS), 
the master neuron fires with $f=63.65$~Hz while the slave-interneuron fires with $f=66.67$~Hz. We believe this mechanism is responsible for the AS phenomenon in different situations. Our conclusion goes in the same direction as that presented by Hayashi and coworkers \cite{Hayashi16} when analyzing systems described by eq. (1).
Additionally, it is worth noting that when $g_{SM}$ is increased in the presence of noise, the spike-timing distribution becomes sharper (Fig.~\ref{fig:tau_gsm}, inset).

\begin{figure}[!ht]%
\begin{flushleft}(a)%
\end{flushleft}%
\vspace{-1.5cm}
\centerline{\includegraphics[width=0.9\columnwidth,clip]{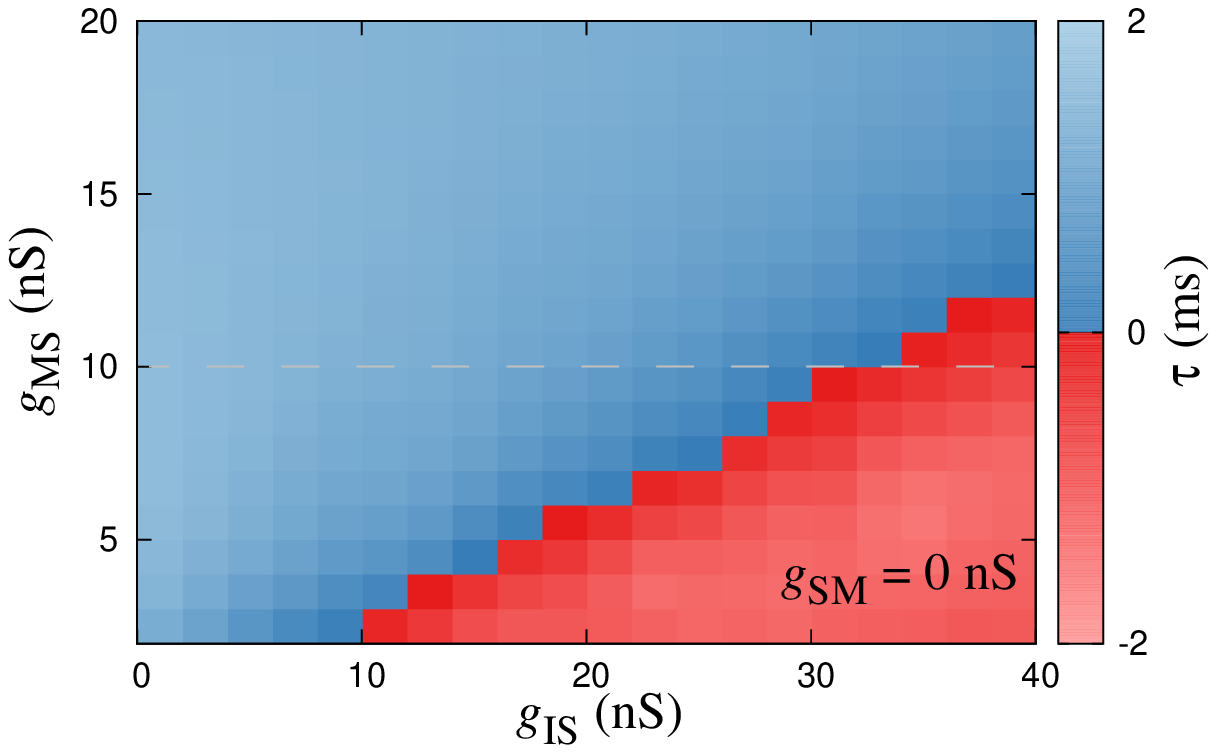}}
\vspace{-0.5cm}
\begin{flushleft}(b)%
\end{flushleft}%
\vspace{-1.5cm}
\centerline{\includegraphics[width=0.9\columnwidth,clip]{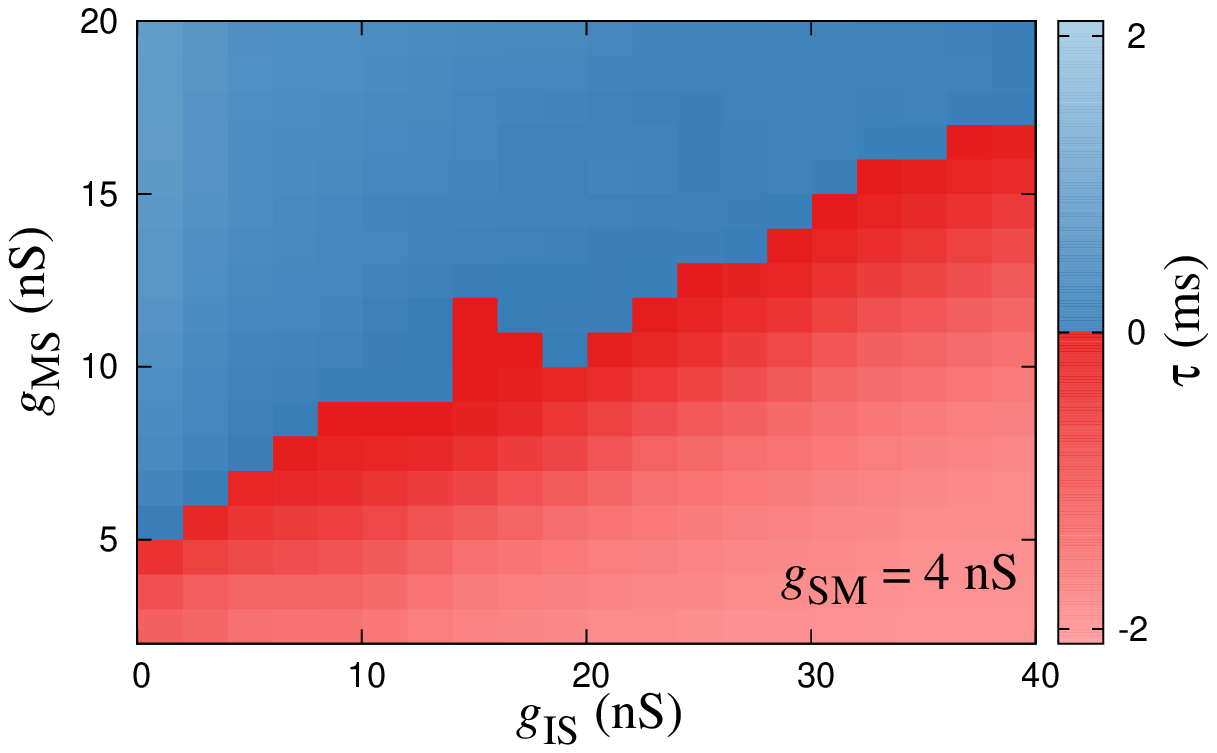}}
\vspace{-0.5cm}
\begin{flushleft}(c)%
\end{flushleft}%
\vspace{-1.5cm}
\centerline{\includegraphics[width=0.9\columnwidth,clip]{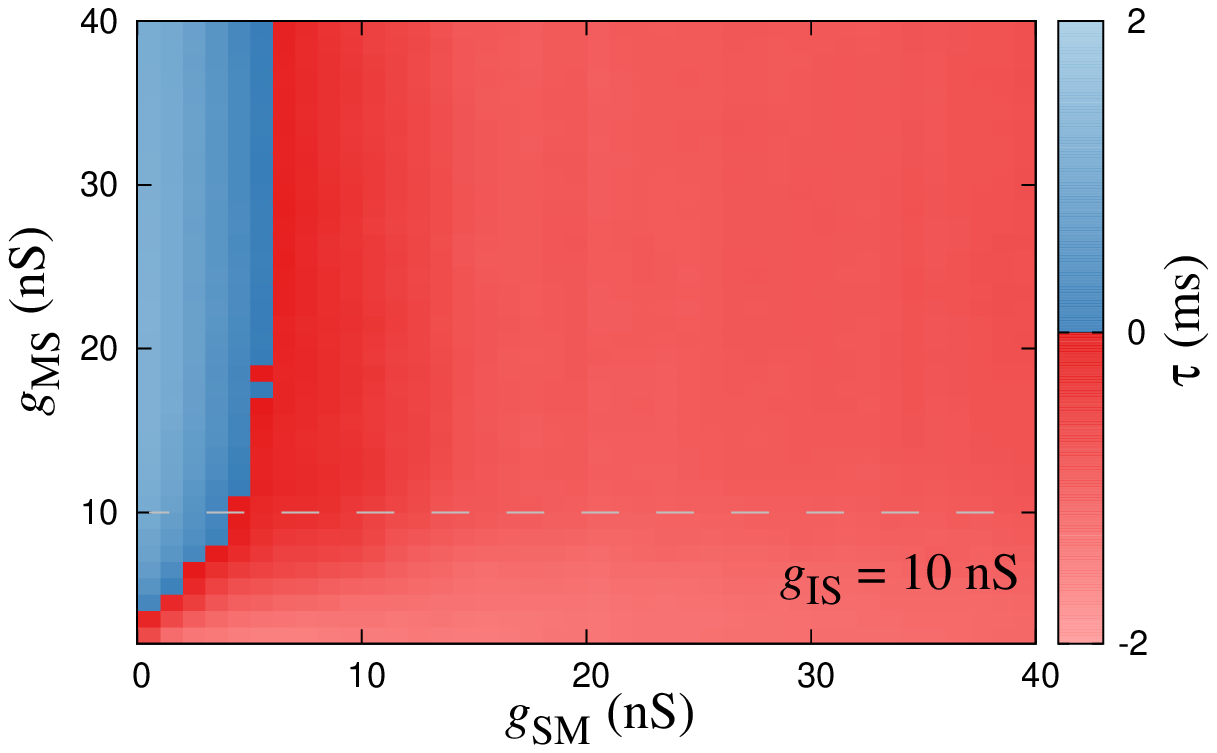}}
\caption{
{\bf (Color online) Dependence of the spike-timing difference with the synaptic conductances.}
$\tau$ is color coded by the right bar in the $(g_{IS},g_{MS})$ projection of parameter space for (a) $g_{SM}=0$ and (b) $g_{SM}=4$~nS.
(c) $\tau$ in the $(g_{SM},g_{MS})$ projection for $g_{IS}=10$~nS.
Dashed lines at $g_{MS}=10$~nS in a and c correspond, respectively, to the curve for $I_c=170$~pA in Fig.~\ref{fig:tau_gis} and to the black curve in Fig.~\ref{fig:tau_gsm}. }
\label{fig:3D} 
\end{figure}

In Fig.~\ref{fig:3D}, we display two-dimensional projections of different
phase diagrams of our model for $I_c=170$~pA. 
We employ the standard values described in Sec.~\ref{model} and Sec.~\ref{Appendix}
except for $g_{MS}$, $g_{SM}$ and $g_{IS}$.
The spike-timing difference $\tau$ is color coded as defined by the right bar.
Negative values of $\tau$ (red) represent regions in which the slave neuron is the leader, 
whereas $\tau>0$ (blue) represents the master leadership. 

In Fig.~\ref{fig:3D}a, $g_{MS}$ and $g_{IS}$ are varied along the vertical and the horizontal axis respectively. 
Since $g_{SM}=0$, negative $\tau$ also accounts for AS, whereas positive $\tau$ means DS.
The two different regimes are distributed in large
continuous regions, with a clear transition between them.
Furthermore, the transition from the DS to the AS regimes can be
well approximated by the linear relation  $g_{MS}/g_{IS} \simeq 0.3$.
Note that the curve represented by circles in Fig.~\ref{fig:tau_gis} corresponds to the horizontal
cut at $g_{MS}=10$~nS in Fig.~\ref{fig:3D}a.

If $g_{MS}=0$ and $g_{SM}\neq0$ the S neuron is the sender and also the leader,
which characterizes the usual DS regime but with the unidirectional connection from the slave to the master. 
In particular, S is the sender and the leader if $g_{SM}>g_{MS}$ and $g_{IS}=0$.
This means that the excitatory synapses from S to M facilitates the leadership of S.
In fact, the presence of the excitatory feedback enlarges the region of $\tau<0$ (compare Fig.~\ref{fig:3D}b, which has $g_{SM}=4$~nS, with Fig.~\ref{fig:3D}a).
For small inhibition, M is the leader if $g_{MS}\gg g_{SM}$, whereas S is the leader if $g_{MS}\ll g_{SM}$.
In Fig.~\ref{fig:3D}c, we show the spike-timing difference $\tau$ 
in the $(g_{SM},g_{MS})$ projection of parameter space for $g_{IS}=10$~nS.
Note that the black curve in Fig.~\ref{fig:tau_gsm} corresponds to the dashed line at $g_{MS}=10$~nS in Fig.~\ref{fig:3D}c.

\section{\label{conclusions}Concluding remarks}

To summarize, we have probed the robustness of the phenomenon of anticipated synchronization in a simple master-slave-interneuron motif of chemically connected neurons. 
In the presence of noise independently applied to the neurons, the spike-timing difference between master and slave neurons $\tau_n$ was shown to have a well defined mean $\tau$. 
We have shown that $\tau$ undergoes a transition from positive to negative values
as a function of the inhibitory synaptic conductance $g_{IS}$, 
corresponding to a transition between delayed and anticipated synchronization. 

Importantly, the distribution of spike-timing difference shows that, regardless of whether the system has $\tau > 0$ (delayed synchronization) or $\tau < 0$ (anticipated synchronization),
there is a non-zero probability that $\tau_n $ eventually exhibits an opposite sign with respect to the mean value $\tau$. 
In practice, this corresponds to a diversity of phase relations that includes changes in sign. 
This is similar to what has been observed experimentally in a variety of setups~\cite{Livingstone96,Maris13,Jia13,Dotson14}.

The phenomenon of anticipated synchronization has also proven to be robust against 
a mutual coupling between master and slave neurons.
Indeed, if AS is present in a unidirectional master-slave connection, 
the non-zero slave-to-master synaptic conductance $g_{SM}$ will not only maintain, but also stabilize the anticipation phenomenon by decreasing its variance. 
For $g_{SM} \geq g_{MS}$, the definitions of master and slave are naturally interchanged. 
In that case, S, the neuron subject to the inhibitory feedback loop (as defined in Fig.~\ref{fig:MSI}), leads M. 
This reinforces the notion that when one excitatory and one inhibitory neuron are mutually connected, 
they may be regarded as a functional unit whose dynamics will typically lead that of another single 
neuron with which it is connected via mutual excitation 
(provided that single neuron does not have an inhibitory loop of itself)~\cite{Gollo14}.

Our results offer a number of possibilities for further investigation.
The interplay between spike-timing dependent plasticity (STDP)~\cite{Song00} and AS can have a major influence
over the structural organization of neuronal networks~\cite{Matias15}. When the 
synchronization regime changes from DS to AS, the mean spike-timing difference between pre and
post-synaptic neurons is inverted, leading to an inversion of the STDP (e.g. from potentiation to
depression). However a study of the combined effects of plasticity, AS and noise in microcircuits is still lacking. 
Interestingly, the 3-neuron motif shown in Fig.~\ref{fig:MSI} can be experimentally reproduced in a hybrid patch
clamp setup as employed by LeMasson et al.~\cite{Masson02}. In such setup the noise plays an important role in the neuronal activity.
Therefore, we believe that our results can be extremely relevant for the verification of AS in vitro.

\section{\label{Appendix}Appendix}

Each neuron in the circuit is represented by a Hodgkin-Huxley model~\cite{HH52}.
It consists of four differential equation 
associating the currents flowing across a patch of an axonal membrane and 
specifying the evolution of the gating variables~\cite{Koch}:
\begin{eqnarray}
C_m \frac{dV}{dt} &=& \overline{G}_{Na} m^3 h (E_{Na}-V) + \overline{G}_{K} n^4 (E_{K}-V) \nonumber \\
&& + G_m (V_{rest}-V) + I_c + I_{P} + \sum I_{syn} \label{eq:dvdt}\\ 
\frac{dx}{dt} &=& \alpha_x(V)(1-x) -\beta_x(V) x \label{eq:alpha} \;.  
\end{eqnarray}
$V$ is the the membrane potential, the ionic currents 
are the Na$^+$, K$^+$ and leakage 
currents, $x\in\{h,m,n\}$ are the gating variables for sodium ($h$ and $m$) and
potassium ($n$). The membrane capacitance of a $30\times 30\times \pi$~$\mu$m$^2$ equipotential patch of
membrane is $C_m = 9\pi$ $\mu$F~\cite{Koch}.
The reversal potentials are $E_{Na}=115$~mV, $E_{K}=-12$~mV and $V_{rest}=10.6$~mV, with maximal conductances $\overline{G}_{Na}=
1080\pi$~mS, $\overline{G}_{K} =324\pi$~mS and $G_m=2.7\pi$~mS,
respectively.
$I_{syn}$ accounts for the chemical synapses from other neurons. 
The voltage-dependent rate constants in the Hodgkin-Huxley model have the form:
\begin{eqnarray}%
\alpha_n(V) & = & \frac{10-V}{100(e^{(10-V)/10}-1)},  \\
\beta_n(V) & = & 0.125e^{-V/80}, \\
\alpha_m(V) & = & \frac{25-V}{10(e^{(25-V)/10}-1)},  \\
\beta_m(V) & = & 4e^{-V/18},\\
\alpha_h(V) & = & 0.07e^{-V/20}, \\
\beta_h(V) & = & \frac{1}{(e^{(30-V)/10}+1)}, \label{eq:betah}
\end{eqnarray}%
where all voltages are measured in mV.

Neurons are coupled through unidirectional excitatory or inhibitory chemical synapses
which, we assume, are mediated by AMPA and GABA$_\text{A}$ receptors, respectively.
The synaptic current received by the post-synaptic neuron is given by:
\begin{equation}
I^{(i)}_{syn} = gr^{(i)}(E_{i}-V),
\label{eq:syn}
\end{equation}
where $V$ is the post-synaptic membrane potential, $g$ the
synaptic conductance and $E_i$ the reversal potential.
The fraction of bound (i.e. open) synaptic receptors $r^{(i)}$ is modeled by a first-order
kinetic dynamics:
\begin{equation}
\label{eq:rate}
\frac{dr^{(i)}}{dt} = \alpha_i [T](1-r^{(i)}) - \beta_i r^{(i)},
\end{equation}
where $\alpha_i$ and $\beta_i$ are rate constants. $[T]$ is the
neurotransmitter concentration in the synaptic cleft.
In its simplest model 
it is an instantaneous function of the pre-synaptic
potential $V_{pre}$~\cite{KochSegev}:
\begin{equation}
[T](V_{pre}) = \frac{T_{max}}{1+e^{[-(V_{pre}-V_p)/K_p]}}.
\end{equation}
In our model $T_{max}=1$~mM$^{-1}$, $K_p=5$~mV, $V_p=62$~mV.
The AMPA and GABA reversal potentials are respectively 
$E_{A}=60$~mV and $E_{G}=-20$~mV.
The rate constants are $\alpha_{A}=1.1$~mM$^{-1}$ms$^{-1}$, $\beta_{A}=0.19$~ms$^{-1}$ ,
$\alpha_{G}=5.0$~mM$^{-1}$ms$^{-1}$, and $\beta_{G}=0.30$~ms$^{-1}$ similarly to the ones in Refs.~\cite{KochSegev,Matias11}.
However, these values depend on a number of different factors and can vary
significantly~\cite{Geiger97,Kraushaar00}.

The Poisson input mimics external excitatory synapses, with conductances
$g_{ext} = 2.0$~nS, from $n$ pre-synaptic neurons,
each one spiking with a Poisson rate $R/n$.
These external excitatory synapses are similar to the AMPA synapses described above, but
$[T]$ was replaced by a Poisson train of quadratic pulses with 1~ms width and 1~mM$^{-1}$ height as shown in Fig.~\ref{fig:Vmemb}a.
The Poisson rate is $R=63$~Hz.
We employed a fourth-order Runge-Kutta algorithm to numerically
integrate the equations with a $0.01$~ms time step.

In order to investigate the existence of the AS phenomenon in neurons spiking with smaller firing rates ($f\simeq17$~Hz)
we have used a modified version of the Hodgkin-Huxley model that includes an extra delayed-rectifier slow K$^{+}$ current to Eq.~\ref{eq:dvdt}~\cite{Pospischil08}:
\begin{equation}
I_{K^+}  =\overline{G}_{M} p (E_{K}-V),
\end{equation}
where $\overline{G}_M=0.07$~mS/cm$^{2}$, $E_{K}=-100$~mV and the gating variable obeys the following equations:
\begin{eqnarray}
\frac{dp}{dt} &=& (p_{\infty}(V) - p)/\tau_{M}(V) \nonumber \\
p_{\infty}(V) &=& \frac{1}{1+e^{-(V+35)/10}}  \nonumber \\
\tau_{M}(V)   &=& \frac{1}{3.3e^{(V+35)/20}+e^{-(V+35)/20}},
\end{eqnarray}
with $\tau_{max}=1$~s. The modified parameters in Eq.~\ref{eq:dvdt} are
$\overline{G}_{Na}=50$~mS/cm$^{2}$,
$\overline{G}_K=5$~mS/cm$^{2}$,
$G_m=0.1$~mS/cm$^{2}$,
$V_{rest}=10.6$~mV,
$E_{K}=-100$~mV
and
$E_{Na}=50$~mV. The voltage-dependent rate constants in Eq.~\ref{eq:alpha} are given by:
 \begin{eqnarray}
 \alpha_n &=& \frac{-0.032(V-V_T-15)}{e^{-(V-V_T-15)/5}-1}  \nonumber \\
\beta_n  &=& 0.5e^{-(V-V_T-10)/40}-1                      \nonumber \\
\alpha_m &=& \frac{-0.32(V-V_T-13)}{e^{-(V-V_T-13)/4}-1}  \nonumber \\
\beta_m  &=& \frac{0.28(V-V_T-40)}{e^{(V-V_T-40)/5}-1}     \nonumber \\
\alpha_h &=&  0.128e^{-(V-V_T-17)/18}                     \nonumber \\
\beta_h  &=& \frac{4}{1+e^{-(V-V_T-40)/5}},                           
 \end{eqnarray} 
with $V_T=55$~mV. Indeed, the resting potential for this model is $-55$~mV.
Synaptic parameters in Eq.~\ref{eq:syn} are modified accordingly to the resting potential: 
$E_{A}=5$~mV and $E_{G}=-75$~mV. The rate constants in Eq.~\ref{eq:rate} are $\alpha_{A}=1.1$~mM$^{-1}$ms$^{-1}$, $\beta_{A}=0.6$~ms$^{-1}$ ,
$\alpha_{G}=5.0$~mM$^{-1}$ms$^{-1}$, and $\beta_{G}=0.60$~ms$^{-1}$. The external applied current is $I_c=900$~pA, the Poissonian rate is $r=9.5$~Hz,
the synaptic conductances are $g_{MS}=g_{SI}=50$~nS, $g_{ext}=0.1$~nS and $g_{SM}=0.0$. For inhibitory conductance $g_{IS}=150$~nS the system presents AS 
(Fig.~\ref{fig:Vmemb}e, $\tau=-1.0$~ms), whereas for $g_{IS}=50$~nS the system exhibts DS.


\begin{acknowledgments}
We thank CNPq grants 480053/2013-8 and 310712/2014-9, FACEPE grant APQ-0826-1.05/15, CAPES grant PVE 88881.068077/2014-01 for financial support. 
\end{acknowledgments}
%

\bibliography{matias}

\end{document}